\documentclass[longauth]{aa}  
\usepackage[varg]{txfonts}

\usepackage{bm} 

\usepackage[T1]{fontenc}
\usepackage{ae,aecompl}
\usepackage{graphicx}   
\usepackage{amsmath}    
\usepackage{amssymb}    

\usepackage{amsmath}	
\usepackage{mathtools}
\usepackage{amssymb}	
\usepackage{natbib}
\bibpunct{(}{)}{;}{a}{}{,} 

\newcommand\ionBrian[2]{#1$\;${\scshape{#2}}}%

\usepackage{hyperref}

\hypersetup{
	colorlinks=true,
	linkcolor=blue,
	citecolor=blue,
	urlcolor=blue,
}

\usepackage{enumerate}
\begin{document}

	\title{
		\centering
		~~~~~~~~~~The ALPINE-ALMA [CII] Survey:
	\newline CGM pollution and gas mixing by tidal stripping in a merging system at $z\sim4.57$}
	\titlerunning{CGM pollution and gas mixing by tidal stripping in a merging system at $z\sim4.57$}
	
	\author{
		M. Ginolfi\inst{\ref{inst1}}
		\and
		G. C. Jones\inst{\ref{inst2},} \inst{\ref{inst3}}
		\and
		M. B\'{e}thermin\inst{\ref{inst4}} 
		\and
		A. Faisst\inst{\ref{inst5}} 
		\and
		B. C. Lemaux\inst{\ref{inst6}} 
		\and
		D. Schaerer\inst{\ref{inst1}}
		\and
		Y. Fudamoto\inst{\ref{inst1}}
		\and ~~~~~~~~~~
		P. Oesch\inst{\ref{inst1}}
		\and
		M. Dessauges-Zavadsky\inst{\ref{inst1}}
	    \and
		S. Fujimoto\inst{\ref{inst7},} \inst{\ref{inst8}}
		\and 
		S. Carniani\inst{\ref{instSNS}} 
	    \and	
		O. Le Fèvre\inst{\ref{inst4}} 
		\and	
		P. Cassata\inst{\ref{inst9}} 
		\and	
		J. D. Silverman\inst{\ref{inst10},} \inst{\ref{inst11}}
		\and	
		P. Capak\inst{\ref{inst5}} 
		\and	
		Lin Yan\inst{\ref{inst12}} 
		\and
		S. Bardelli\inst{\ref{inst13}} 
		\and
		O. Cucciati\inst{\ref{inst13}} 
		\and
		R. Gal\inst{\ref{inst15}} 
		\and
		C. Gruppioni\inst{\ref{inst13}} 
		\and
		N. P. Hathi\inst{\ref{inst16}}
		\and ~~~
		L. Lubin\inst{\ref{inst6}} 
		\and
		R. Maiolino\inst{\ref{inst2},} \inst{\ref{inst3}}
		\and
		L. Morselli\inst{\ref{inst9}} 
		\and
		D. Pelliccia\inst{\ref{inst6}} 
		\and
		M. Talia\inst{\ref{inst13},} \inst{\ref{inst17}}
		\and
		D. Vergani\inst{\ref{inst13}} 
		\and
		 G. Zamorani\inst{\ref{inst13}} 
	}
	
	\institute{
		Observatoire de Gen\`eve, Universit\`e de Gen\`eve, 51 Ch. des Maillettes, 1290 Versoix, Switzerland\\
		\email{michele.ginolfi@unige.ch}\label{inst1}
		\and
		Cavendish Laboratory, University of Cambridge, 19 J. J. Thomson Ave., Cambridge CB3 0HE, UK\label{inst2}
		\and
		Kavli Institute for Cosmology, University of Cambridge, Madingley Road, Cambridge CB3 0HA, UK\label{inst3}
		\and
		Aix Marseille Univ, CNRS, CNES, LAM, Marseille, France\label{inst4}
		\and
		IPAC, California Institute of Technology, 1200 East California Boulevard, Pasadena, CA 91125, USA\label{inst5}
		\and
		Department of Physics, University of California, Davis, One Shields Ave., Davis, CA 95616, USA\label{inst6}
		\and
		Cosmic Dawn Center (DAWN) \label{inst7}
		\and
		Niels Bohr Institute, University of Copenhagen, Lyngbyvej 2, DK-2100 Copenhagen, Denmark \label{inst8}
		\and
		Scuola Normale Superiore, Piazza dei Cavalieri 7, Pisa IT-56126, Italy\label{instSNS}
		\and
		University of Padova, Department of Physics and Astronomy Vicolo Osservatorio 3, 35122, Padova, Italy\label{inst9}
		\and
		Department of Astronomy, School of Science, The University of Tokyo, 7-3-1 Hongo, Bunkyo, Tokyo 113-0033, Japan\label{inst10}
		\and
		Kavli Institute for the Physics and Mathematics of the Universe, The University of Tokyo, Kashiwa, Japan 277-8583 (Kavli IPMU, WPI)\label{inst11}
		\and
		Caltech Optical Observatories, Cahill Center for Astronomy and Astrophysics 1200 East California Boulevard, Pasadena, CA 91125, USA\label{inst12}
		\and
		 Istituto Nazionale di Astrofisica - Osservatorio di Astrofisica e Scienza dello Spazio, via Gobetti 93/3, I-40129, Bologna, Italy\label{inst13}
		\and		
		University of Hawai’i, Institute for Astronomy, 2680 Woodlawn Drive, Honolulu, HI 96822, USA\label{inst15}
		\and
		Space Telescope Science Institute, 3700 San Martin Drive, Baltimore 21218, USA\label{inst16}
		\and	
		University of Bologna, Department of Physics and Astronomy (DIFA), Via Gobetti 93/2, I-40129, Bologna, Italy\label{inst17}		
	}
	
	\date{Received XXX; accepted YYY}

	\abstract{
		We present ALMA observations of a merging system at $z\sim4.57$, 
		observed as a part of the \textit{ALMA Large Program to INvestigate [CII] at Early times} (ALPINE) survey.
		Combining ALMA [CII]158$\,\mu$m and far-infrared continuum data with
		multi-wavelength ancillary data we find that the system is composed of two massive ($M_{\star} \gtrsim 10^{10} ~ {\rm M_\odot}$) star-forming galaxies experiencing a major merger (stellar mass ratio $r_{\rm mass} \gtrsim 0.9$) at close spatial ($\sim 13$ kpc; projected) and velocity ($\Delta v < 300 ~ {\rm km~s^{-1}}$) separations, 
		and two additional faint narrow [CII]-emitting satellites.
		The overall system belongs to a larger-scale protocluster environment and is coincident to one of its overdensity peaks.
		ALMA reveals also the presence of
		 [CII] emission 
		arising from a circumgalactic gas structure, extending up to a diameter-scale of $\sim 30$ kpc.
		Our morpho-spectral decomposition analysis shows that
		about $50 \%$ of the total flux resides {\it between} the individual galaxy components, 
		in a metal-enriched gaseous envelope characterized by a disturbed morphology and complex kinematics.
		Similarly to observations of shock-excited [CII] emitted from tidal tails in local groups,
		our results can be interpreted as a possible signature of interstellar gas stripped by strong gravitational interactions,
		with a possible contribution from material ejected by galactic outflows and emission triggered by star formation in small faint satellites.
		Our findings suggest that mergers could be an efficient mechanism of gas mixing in the circumgalactic medium around high-$z$ galaxies,
		and thus play a key role in the galaxy baryon cycle at early epochs.
		}
	
	
	\maketitle

\section{Introduction}\label{sec:introduction}

Modern theories of galaxy evolution and cosmological numerical simulations predict that high-$z$ galaxies assembled their mass through a combination of cold gas accretion from the intergalactic/circumgalactic medium (IGM/CGM) and merging activity
(e.g., \citealp{Bower2006, Dekel2009, Hopkins2010, Vogelsberger2014, Schaye2015}).
%
Continuous baryonic flows are needed to replenish the gas content, fuel the star formation and drive the evolution of galaxies on the  star-forming Main Sequence (e.g., \citealp{Daddi2010,Rodighiero2011, Lilly2013, Speagle2014, Scoville2016}).

On the other hand strong dynamical interactions between galaxies, such as major mergers, can efficiently drive a significant amount of gas into the central regions (e.g., \citealp{Barnes1991}) and both 
(i) boost the efficiency of star formation (see \citealp{Elmegreen2000}), triggering starbursts (e.g., \citealp{Sanders1996, Hopkins2006, DiMatteo2008, Bournaud2011}), 
and (ii) feed the growth of super massive black holes (SMBHs), thereby powering Active Galactic Nuclei (AGN) activity (e.g., \citealp{GarciaBurillo2005, DiMatteo2005, Ginolfi2019}).
Moreover, mergers can significantly disturb and shape the morphologies of the galaxies involved, leading to the formation of tidal tails.
Similarly to the ram pressure stripping  (\citealp{Gunn1972}) extensively observed in nearby galaxy clusters (e.g., \citealp{Boselli2006, Moretti2018, Poggianti2019, Fossati2019}),
tidal tails in merging systems are made of gas stripped from outer regions of galaxies, and extend well  beyond the disks, on circumgalactic scales (e.g., \citealp{Bridge2010, Wen2016, Guo2016, Stewart2011}).
As a result, strong dynamical interactions between galaxies contribute to the intergalactic transfer of processed material as shown by recent simulations 
(e.g., \citealp{Nelson2015, Angles2017, Hani2018}), 
and, at the same time, to the chemical enrichment of the CGM (e.g., \citealp{Bournaud2011, Graziani2020}).


While tidal tails are well studied at different wavelengths in the local and in the intermediate redshift Universe (e.g., \citealp{Poggianti2017, Vulcani2017}),
similar observations become challenging at higher redshift, mainly because of the diffuse and faint nature of the stripped gas.
%
%
A solution comes from the Atacama Large Millimeter/submillimeter Array (ALMA),
which enables us to possibly study the efficiency of tidal gas stripping and circumgalactic gas mixing in the Early Universe, 
by mapping the morphology and the kinematics of the CGM around merging systems, using bright far-infrared (FIR) lines, such as [CII] 158 $\mu$m (hereafter [CII]; see \citealp{Carilli2013}).
[CII] has been recently detected in several star-forming galaxies at $z>4$ 
(e.g., \citealp{Capak2015, Inoue2016, Pentericci2016, Bradac2017, Carniani2018, Matthee2019, Schaerer2020, Carniani2020}), 
some of which are experiencing major merger events (e.g., \citealp{Oteo2016, Riechers2017, Pavesi2018, Marrone2018, Decarli2019, Jones2020}),
or less disruptive gravitational interactions, usually revealed in combination with clumpy morphologies (e.g., \citealp{Maiolino2015, Matthee2017,Jones2017, Carniani2018b}).
[CII] is generally emitted from multiple phases (ionized, neutral and molecular gas) of the interstellar medium 
(ISM; e.g., \citealp{Vallini2013, Vallini2017, Lagache2018, Ferrara2019}), 
and it has been shown to be a good tracer of shock-induced emission in the surroundings of local interacting galaxies (e.g., \citealp{Cormier2012,Appleton2013,Velusamy2014}).

In this paper we present a study of the morpho-kinematic properties of [CII] emission in/around a major merging system at $z\sim4.57$, observed as a part of the {\it ALMA Large Program to Investigate C$^+$ at Early Times} (ALPINE) survey, 
that measured [CII] and FIR-continuum emission in a statistical sample of more than a hundred normal galaxies with spectroscopic redshifts between $4<z<6$ (see \citealp{LeFevre2020}, \citealp{Bethermin2020} and \citealp{Faisst2019} for descriptions of (i) the survey objectives, (ii) the ALMA data-processing, and (iii) the multi-wavelength ancillary dataset, respectively).

The paper is organized as follows.
In Sec. \ref{sec:Observations} we describe the ALMA data reduction process, in Sec. \ref{sec:Results} we report our results, and in Sec. \ref{sec:discussion} we discuss their implications. Conclusions are summarized in Sec. \ref{sec:conclusions}. 
Throughout the paper we assume a flat $\Lambda$CDM cosmology ($\Omega_\Lambda = 0.7$, $\Omega_m = 0.3$, $H_0 = 70$ km s$^{-1}$ Mpc$^{-1}$), and adopt a Chabrier initial mass function (IMF; \citealp{Chabrier2003}).
At the redshift of our target vuds\_cosmos\_5101209780 ($\rm VC\_9780$;
$z_{\rm [CII]}$ = 4.573), 1 arcsecond corresponds to 6.55 proper kpc.

\section{ALMA Observations}\label{sec:Observations}

%
ALMA observations of the merging system studied in this work were conducted in Band 7 on 2 June 2018, during Cycle 5 (project 2017.1.00428.L, PI O. Le Fèvre), using a C43-2 array configuration with 45 antennae, and a total on-source time of 20 minutes.
The spectral setup consisted of two sidebands, 
each composed of two spectral windows with total bandwidths of 1.875 GHz and channel width resolutions of 15.6 MHz.
The lower sideband was tuned to the expected [CII] redshifted frequency, according to the spectroscopic redshift extracted from the rest-frame UV spectrum (drawn from the VUDS survey; \citealp{LeFevre2015, Tasca2017}).
The upper sideband is used to search for FIR-continuum emission.
The phase was centered at the rest-frame UV position of $\rm VC\_9780$ (see Sec. \ref{sec:ancillary}; $\rm RA: 10h01m33.45s$; $\rm DEC:  +02d22m10.19s$).
The data were calibrated using the automatic pipeline provided by the Common Astronomy Software Applications package (\texttt{CASA}; \citealp{McMullin2007}), version 5.4.0.
No additional flagging was needed as the pipeline calibration output did not show any issue.
The continuum map was obtained running the \texttt{CASA} task \texttt{tclean} (multi-frequency synthesis mode) over the line-free visibilities in all spectral
windows.
The [CII] datacube was generated from the continuum-subtracted visibilities (CASA task \texttt{uvcontsub}), using the following setting:  
500 iterations and a signal-to-noise ratio (S/N) threshold of $\sigma_{\rm clean} = 3$ in the cleaning algorithm, a pixel size of 0.15$''$ and a spectral binning of 25 km s$^{-1}$ ($\sim 29$ MHz).
For both continuum and [CII] datacube we adopted a natural weighting of the visibilities to maximise the sensitivity.
We reach an average root mean square (RMS) noise level at the phase center of $\sigma_{\rm cube} \sim 14$ mJy beam$^{-1}$ 
for a 25 km s$^{-1}$ spectral channel in the [CII] datacube, and of $\sigma_{\rm cont} \sim 60$ $\mu$Jy beam$^{-1}$  in the continuum image.
The synthesized beam in our final data-products is $1.21''  \times 0.77"$, with a position angle of $-61^{\circ}$.
Additional information on the overall ALPINE data-processing strategy (including data quality assessment) can be found in \cite{Bethermin2020}.

\section{Analysis and Results}\label{sec:Results}

\begin{figure*}
	\centering
	\includegraphics[width=1\textwidth]{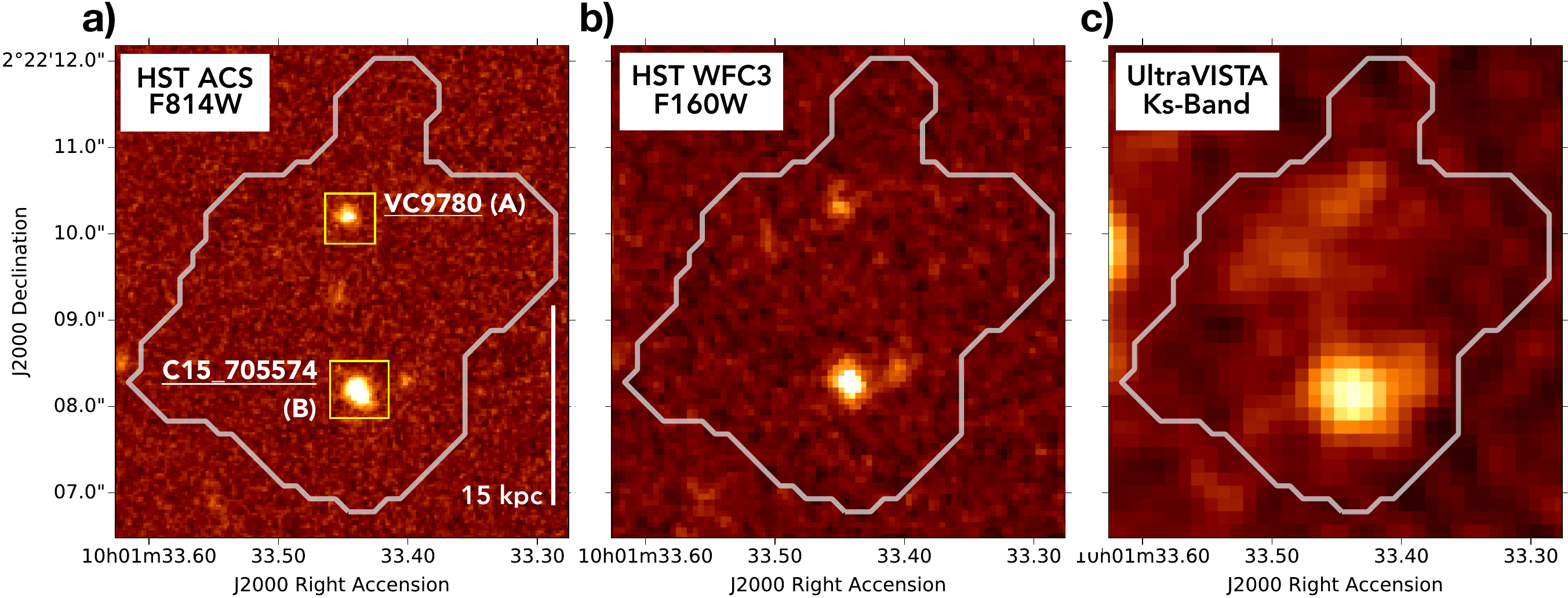}
	\caption{
		{\it (a)} HST/ACS F814W (i-band; \citealp{Koekemoer2007}), {\it (b)} HST/WFC3 F160W (Faisst et al., in prep.), and {\it (c)} UltraVISTA DR4 Ks-band (\citealp{McCracken2012}) images of the system.
		The rest-frame UV positions of $\rm VC\_9780$ (dubbed $\#A$) and  $\rm C15\_705574$ (dubbed $\#B$) are highlighted with a yellow square in {\it (a)}.
		The grey contour indicates the extension of the full [CII]-emitting region (see Sec. \ref{sec:almaAnalysis}).
		The white bar shows a reference length of 15 kpc. North is up and east is to the left.
		}
	\label{fig:fig1}
\end{figure*}

\begin{figure}
	\centering
	\includegraphics[width=1\columnwidth]{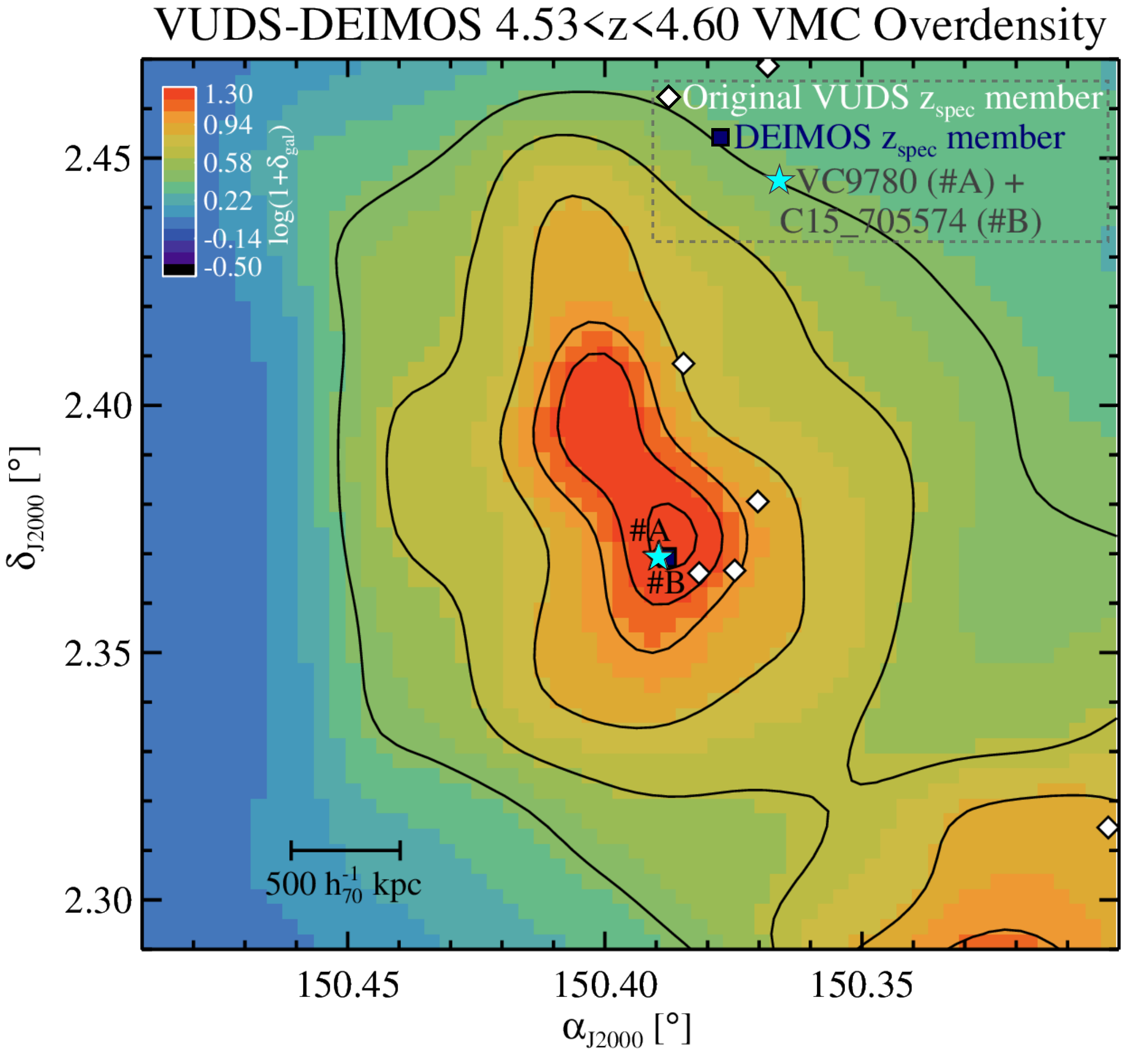}
	\caption{
VMC galaxy overdensity map of the N-E region of the ${\rm PCl J1001+0220}$ proto-cluster (see details in Sec. \ref{sec:protocluster}).
The VUDS spectral members of the system are indicated with white diamond symbols, while the location of our merging system and a possibly associated additional member observed with DEIMOS are indicated with a cyan star and a blue square, respectively.
Over-plotted on the map are overdensity isopleths at significance levels of $[4, 6, 8, 10, 11.5, 13]\sigma$, where $\sigma$ is calculated as the RMS fluctuation of the overdensity field over the entire map (see \citealp{cucciati2018, hung2020}). 
The map shows that the merging system studied in this work is clearly located close to an overdensity peak of the proto-cluster.
The black solid bar shows a reference length of 500 $h_{70}^{-1}$ kpc.
	}
	\label{fig:fig2}
\end{figure}

\begin{figure*}
	\centering
	\includegraphics[width=1\textwidth]{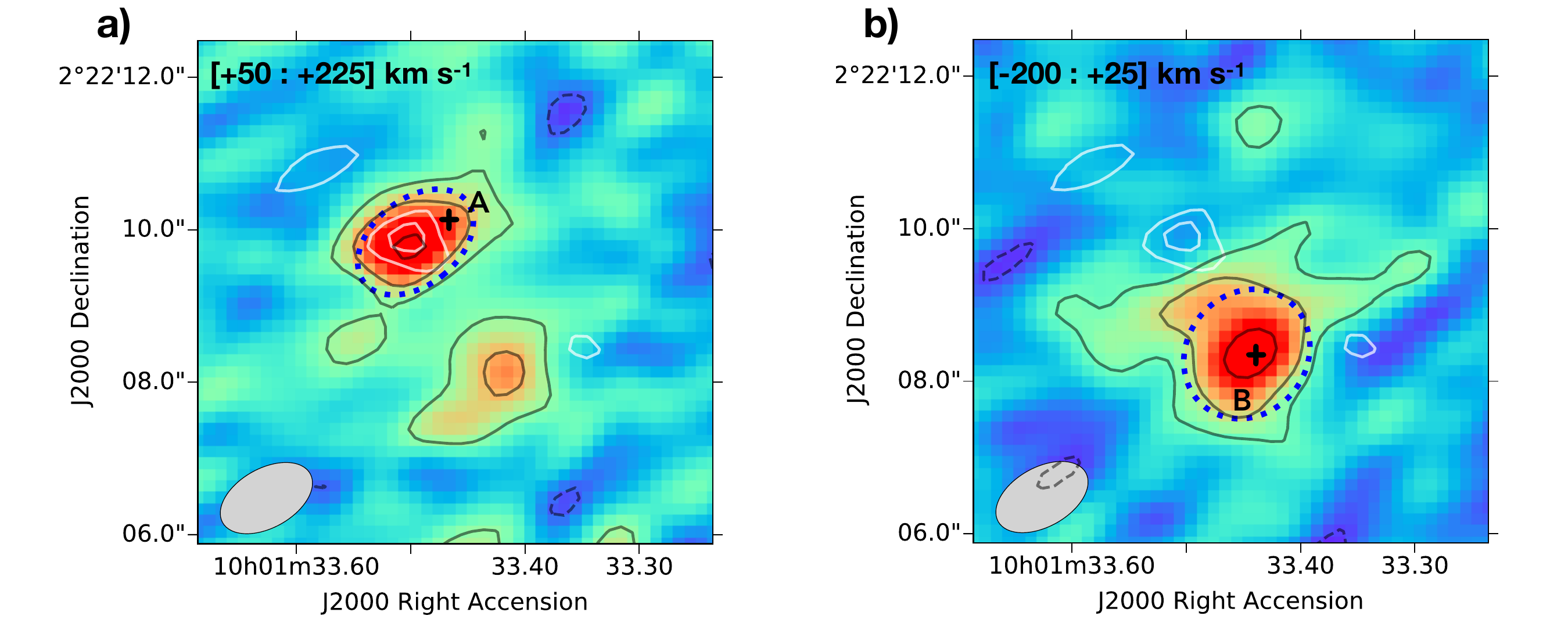}
	\caption{
		{\it (a)} Velocity-integrated [CII] map of $\rm VC\_9780$ ($\#A$) and  {\it (b)} $\rm C15\_705574$ ($\#B$), 
		obtained by collapsing channels (velocity ranges in the upper-left corners) of line-emission in 1$''$-aperture spectra (see Sec. \ref{sec:almaAnalysis}) centered on the rest-frame UV centroids of $\#A$ and $\#B$ (indicated with a black plus symbol).
		Black solid (dashed) contours indicate the positive (negative) significance levels at $[2, 4, 8]  \sigma$ of [CII] flux, where 
		$\sigma_{\rm[CII]} =  50$ mJy km s$^{-1}$ beam$^{-1}$ in  {\it (a)} and $\sigma_{\rm[CII]} =  57$ mJy km s$^{-1}$ beam$^{-1}$ in  {\it (b)}.
	    The white contours indicate the positive significance levels at $[3, 4]  \sigma$ of FIR-continuum emission, where 
		$\sigma_{\rm cont} =  59$ $\mu$Jy/beam. No negative contours at the same significance levels were found.
		The dashed blue ellipses indicate the beam-convolved ${\rm FWHM}_x \times {\rm FWHM}_y$ regions obtained by 2D Gaussian models, 
		and correspond to the apertures used to extract the final [CII] spectra of $\#A$ and $\#B$ (see Sec. \ref{sec:almaAnalysis}), shown in Fig. \ref{fig:fig4}.
		The ALMA beam size is given in the bottom-left corners. North is up and east is to the left.
	}
	\label{fig:fig3}
\end{figure*}

\subsection{A major merging system at $z\sim 4.57$}\label{sec:ancillary}

Using the available ancillary data and catalogues we first identify the system studied in this paper to be composed of two massive and rest-frame UV bright galaxies (see an HST/ACS F814W cutout in Fig. \ref{fig:fig1}a), at a mutual projected distance of $\sim$ 13 kpc:
%
%
(1) $\rm VC\_9780$, a Lyman-break galaxy at $z_{\rm UV} = 4.57$ (confirmed $z_{\rm [CII]} = 4.573$, by ALPINE), 
with a stellar mass of $M_{\star} = 1.1^{+0.4}_{-0.3} \times 10^{10} ~ {\rm M_\odot}$
and a star formation rate of ${\rm SFR} = 38^{+29}_{-14} ~ {\rm M_\odot ~ yr^{-1}}$ (see \citealp{Faisst2019});
%
%
(2) $\rm C15\_705574$%
\footnote{$ID=705574$ in the  COSMOS2015 catalogue (\citealp{Laigle2016}).}%
, at a photometric redshift $z_{\rm phot} = 4.62$ (confirmed $z_{\rm [CII]} = 4.568$, by ALPINE), 
with  $M_{\star} = 1.2^{+1.0}_{-0.2} \times 10^{10} ~ {\rm M_\odot}$
and ${\rm SFR} = 106^{+9}_{-65} ~ {\rm M_\odot ~ yr^{-1}}$ (from the COSMOS2015 catalogue%
\footnote{The physical properties of $\rm C15\_705574$ reported in the COSMOS2015 catalogue are computed adopting the photometric redshift. However, given the high accuracy of the latter ($|z_{\rm phot} - z_{\rm [CII]}| \sim0.05$; see text) we assume $M_{\star}$ and  ${\rm SFR}$ to be correct within their uncertainties.
}%
; \citealp{Laigle2016}).     
To simplify the reading, we dub $\rm VC\_9780$ and $\rm C15\_705574$ as $\#A$ and $\#B$, respectively.
%
%

For both objects, $M_{\star}$ and SFR are measured through the spectral energy distribution (SED) modelling code  \texttt{LePHARE} (\citealp{Arnouts1999, Ilbert2006, Arnouts2011}) ran over the broadband photometry (including ground- and space-based imaging from the rest-frame UV to the near-IR; see \citealp{Faisst2019}) available for the well-studied \textit{COSMOS} field (\citealp{Scoville2007}).
%
From the HST/ACS F814W image, modelling the light profiles with a Sérsic function (\citealp{Sersic1963}) and assuming a Sérsic index n=1 (i.e., an exponential-disk profile), we estimate rest-frame UV effective radii of $r_{\rm e,UV} = 1.0 \pm 0.24$ kpc and $r_{\rm e,UV} = 1.1 \pm 0.3$ kpc for $\#A$ and $\#B$, respectively (see \citealp{Fujimoto2020} for an accurate description of the size-fitting procedure for ALPINE galaxies).

In Fig. \ref{fig:fig1}b we show a HST/WFC3 F160W cutout
of the system%
\footnote{Data drawn from a recently approved mid-cycle-26 HST program (Faisst et al., in prep.).} %
(rest-frame $\lambda_{\rm rest}\sim2800 ~\AA$),
while in Fig. \ref{fig:fig1}c we display the 
UltraVista Ks-Band image from the fourth data release 
($\rm DR4$%
\footnote{\url{http://www.eso.org/rm/api/v1/public/releaseDescriptions/132}}%
; rest-frame $\lambda_{\rm rest}\sim4000 ~\AA$), showing a faint elongation of $\#A$ towards the South-East (S-E; see also Sec. \ref{sec:almaAnalysis}).

The close separation both in spatial projected distance ($\sim 13$ kpc in the HST rest-frame UV images) and redshift ($|\Delta z| = 0.004$, corresponding to an absolute velocity offset of only 270 km s$^{-1}$; see Sec. \ref{sec:almaAnalysis}) suggests that $\#A$ and $\#B$ are gravitationally bound.
Moreover, the ratio of their stellar masses, $r_{\rm mass}  \sim 0.9$,  is very close to unity, indicating that the galaxy system is undergoing a close major merger event (e.g., \citealp{Stewart2009, Lotz2010, Jones2020}).


\subsection{Location at the density peak of a protocluster environment}\label{sec:protocluster}

In order to explore the larger context in which this merging event is occurring, we relied on the wealth of other spectroscopic and photometric data available in the COSMOS field to map out the large-scale galaxy density surrounding our system. 
The mapping was done using the Voronoi Monte Carlo (VMC) technique, 
first presented in \cite{tomczak2017}, \cite{lemaux2017,lemaux2019}, and \cite{hung2020} for use at $z\sim1$,
and adapted for studies at higher redshift ($2 \leq z \leq 5$) in \cite{cucciati2018} and \cite{lemaux2018}. 
This technique statistically combines both spectroscopic and photometric redshifts to generate a suite of overlapping two-dimensional density maps in fine redshift slices over an arbitrary redshift range, and it has been shown to have high fidelity, precision, and accuracy in reconstructing the galaxy density field at a variety of different redshifts and for a variety  of different observational data (see \citealp{lemaux2018};  Lemaux et al., in prep.).
While we refer to the works cited above for more details, we briefly describe the method in the Appendix \ref{sec:appendix1}.

Our system is located in close projected spatial and redshift proximity to the massive proto-cluster 
${\rm PCl J1001+0220}$ at $z\sim4.57$, discovered by \cite{lemaux2018}.
Thus, we focus here on generating a map that spans the fiducial redshift boundaries of ${\rm PCl J1001+0220}$ ($4.53 \le z \le 4.60$). 
Such a redshift range corresponds to 7.5 $h_{70}^{-1}$ proper kpc, or $\sim3750$ km s$^{-1}$. 
As in \cite{lemaux2018}, we use the full VUDS and zCOSMOS 
(\citealp{lilly2007, lilly2009, diener2013, diener2015}; Lilly et al., in prep.) 
catalogs as our primary spectroscopic catalogs, 
with photometric redshifts derived from fitting to the photometry presented in the COSMOS2015 catalog (\citealp{Laigle2016}). 
In addition to the VUDS and zCOSMOS catalogs, we incorporate here new \ionBrian{Keck}{ii}/Deep Imaging Multi-Object Spectrograph (DEIMOS, \citealp{faber2003}) 
observations of ${\rm PCl J1001+0220}$ (see a description of the observations in Lemaux et al., in prep.)
which resulted in the spectroscopic confirmation of 106 new galaxies in the region in and around ${\rm PCl J1001+0220}$.
%
%
%

Shown in Fig. \ref{fig:fig2} is the resultant VMC galaxy overdensity map for the north-east (N-E) region of the ${\rm PCl J1001+0220}$ proto-cluster.
Also shown are the original VUDS spectral members of the system (see \citealp{lemaux2018} for more details), as well as 
our $\#A$  --  $\#B$ system. 
One of the new members confirmed by the DEIMOS observations ($ID= 705080$) is also shown. 
This galaxy, 
with a $M_{\star} = 5.3^{+2.5}_{-2.1}\times10^9 ~{\rm M_\odot} $
and a ${\rm SFR} = 22.2^{+17.1}_{-8.7} ~ {\rm M_\odot ~ yr^{-1}}$,
is located at $\sim35$ $h_{70}^{-1}$ proper kpc from our merging system, 
although it is not detected by ALMA, neither in [CII] nor in the rest-frame FIR continuum.
Measuring its Ly$\alpha$ we determine a redshift $z_{Ly\alpha}=4.5769$, corresponding to a velocity offset of $\Delta v \sim 260 ~ {\rm km~s^{-1}}$ with respect to $\#A$ and $\Delta v \sim 580 ~ {\rm km~s^{-1}}$ with respect to $\#B$%
\footnote{We note that the actual velocity offsets are likely to be even lower, since the redshift measured through Ly$\alpha$ can be off by few hundreds of $ {\rm km~s^{-1}}$ toward the red (see e.g., \citealp{Shapley2003, Trainor2015, Verhamme2018, Cassata2020}).}. 
Such a close projected spatial and redshift proximity to 
$\#A$  --  $\#B$
suggests that DEIMOS $ID= 705080$ is possibly associated to them.

As can be seen in  Fig. \ref{fig:fig2}, our merging system at $z\sim4.57$ is nearly coincident with the overdensity peak of the N-E sub-component of the proto-cluster ${\rm PCl J1001+0220}$, and lie within $\leq$ 385 pkpc from the barycenter of this sub-component.
The average overdensity across this region is $\log(1+\langle \delta_{gal}\rangle)=1.5$, comparable to the cores of $z\sim1$ groups or the outskirts of massive clusters (e.g., \citealp{lemaux2017, tomczak2017}). 
Altogether these findings indicate that the merging system presented in this paper belongs to a large-scale dense and complex proto-cluster environment. 
In Sec. \ref{sec:discussion} we discuss possible implications of this scenario in the interpretation of our results.


In the next section we describe the results of ALMA observations of our system.

\begin{figure*}
	\centering
	\includegraphics[width=1\textwidth]{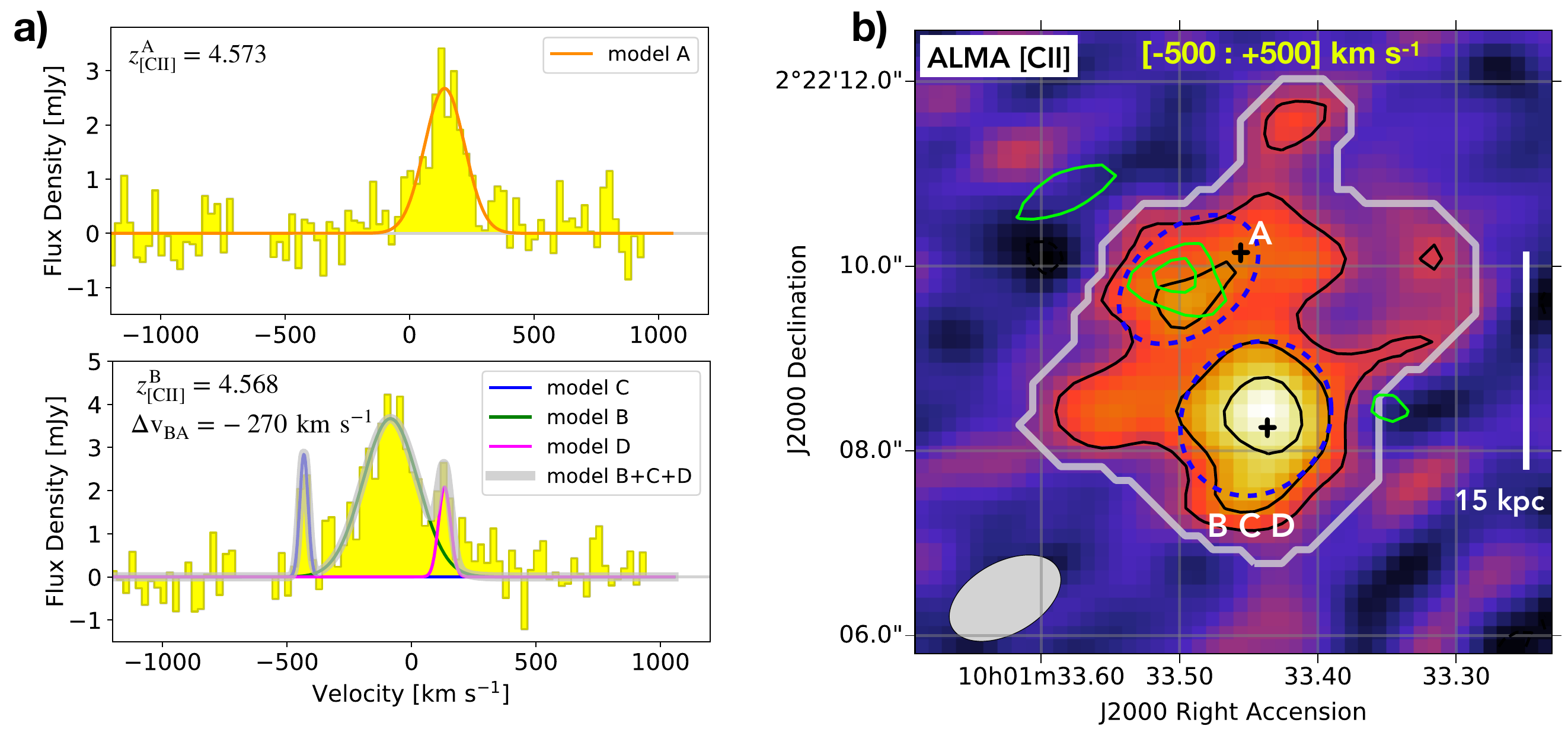}
	\caption{
		{\it (a)}
		[CII] spectra  (yellow) of the individual galaxy components in the merging system are shown: $\#A$ in the upper panel, and the combined spectrum of $\#B$, $\#C$, $\#D$ in the lower panel. Solid coloured lines represent Gaussian models of the spectra.
		{\it (b)}
		Velocity-integrated [CII] flux map obtained integrating the broad velocity range $\rm [-500 : +500] ~ {\rm km~s^{-1}}$,
		to visualize the total [CII] emission arising from the full system. 
		Black plus symbols show the rest-frame UV centroids of $\#A$ and $\#B$.	
		The dashed blue ellipses defined in Fig. \ref{fig:fig3} (see Sec. \ref{sec:almaAnalysis}) are shown for reference.
		Black solid (dashed) contours indicate the positive (negative) significance levels at $[2, 4, 6] \sigma$ of [CII] flux, where 
		$\sigma_{\rm[CII]} =  0.13$ Jy km s$^{-1}$ beam$^{-1}$.
		Green contours indicate (as in Fig. \ref{fig:fig3}) the FIR-continuum significance levels $[3, 4]  \sigma$, where 
		$\sigma_{\rm cont} =  59$ $\mu$Jy/beam.
		The grey solid line indicates the 1$\sigma$ level of the total [CII] emission, and it is used as a reference aperture for the extraction of the full [CII] spectrum arising from the entire system.
		The ALMA beam size is given in the bottom-left corner. A scale of 15 kpc is shown in the right side. North is up and east is to the left.
	}
	\label{fig:fig4}
\end{figure*}

\subsection{ALMA morpho-kinematics analysis}\label{sec:almaAnalysis}

\begin{table*}\label{tab:table}
	\centering
	\begin{tabular}{llllll}
		\hline
		~  & ~& ~ & ~ & ~ & ~\\ [-1.5ex]
		~ & $M_{\star}$ [${\rm M_\odot}$] & ${\rm SFR}$ [${\rm M_\odot ~ yr^{-1}}$] & $L_{\rm [CII]}$ [${\rm L_\odot}$] & ${\rm FWHM}_{\rm [CII]}$  [km s$^{-1}$]& $z_{\rm [CII]}$ \\  [+0.5ex]
		\hline 
		~  & ~& ~ & ~& ~ & ~\\ [-1.5ex]
		$\rm VC\_9780$ ($\#A$) & $1.1^{+0.4}_{-0.3} \times 10^{10}$ & $38^{+29}_{-14}$ & $3.5 ~(\pm 0.3) \times 10^8$ & $190$ & 4.573	 \\ [+0.5ex]
		$\rm C15\_705574$ ($\#B$) & $1.2^{+1.0}_{-0.2} \times 10^{10}$ & $106^{+9}_{-65}$ & $6.6 ~(\pm 0.4) \times 10^8$ & $260$ & 4.568 \\ [+0.5ex]
		$\#C$ & -- &  --  & $7.7 ~\pm 1.5 \times 10^7$ & 40 & 4.562 \\ [+0.5ex]
		$\#D$ & -- &  --  & $8.4 ~\pm 1.9 \times 10^7$ & 57 & 4.573 \\ [+1ex]
		\hline
		~  & ~& ~ & ~& ~& ~\\
	\end{tabular}
	\caption{$M_{\star}$ and SFRs (from SED-fitting; see Sec. \ref{sec:ancillary}), 
		along with  $L_{\rm [CII]}$, ${\rm FWHM}_{\rm [CII]}$ and $z_{\rm [CII]}$ (measured with ALMA; see Sec. \ref{sec:almaAnalysis}),
		are summarized for galaxies and satellites in the merging system.
	}
\end{table*}

We use our ALPINE ALMA observations (see Sec. \ref{sec:Observations}) of the [CII] line emission to perform a three-dimensional (3D) morpho-spectral decomposition of our merging system.

We start by extracting two spectra from large 
1$''$ (diameter)-apertures centered on the rest-frame UV centroids of $\#A$ and $\#B$.

We then produce velocity-integrated maps of the [CII] emission in/around $\#A$ and $\#B$ (see Fig. \ref{fig:fig3}) by collapsing the cube-channels included in the spectral range [$f_{\rm cen} - \sigma_{\rm G}$, $f_{\rm cen} + \sigma_{\rm G}$], where $f_{\rm cen}$ and $\sigma_{\rm G}$ are the central 
frequency 
and the standard deviation from a Gaussian model of the spectra. 
The corresponding velocity ranges are labelled in the top-left corners of both panels of Fig. \ref{fig:fig3}, defined adopting a zero-velocity reference at $z_{\rm ref} = 4.570$ (see next paragraphs of this Section).
Although this will be discussed in detail further on by means of tailored analyses, we note that maps in Fig. \ref{fig:fig3} already suggest a disturbed morphology of the system, showing the presence of complex resolved extended emission around the sources.
%
%

We then estimate the spatial extent of the [CII]-emission arising from $\#A$ and $\#B$ by fitting a simple 2D Gaussian model to the velocity-integrated maps (Fig. \ref{fig:fig3}), and use the ${\rm FWHM}_x \times {\rm FWHM}_y$ ellipses as apertures to extract the correct spectral information%
\footnote{As an alternative spectra-extraction procedure, we tested a 1$''$-aperture centred on the [CII] peaks, and the results are virtually unchanged, not affecting our interpretation.}.

This three-step process is needed to allow for any peculiar geometry or spatial offset between the [CII] and the rest-frame UV emission. 
Indeed, we note that the [CII] peak of $\#A$ is offset by $\gtrsim 0.5''$ from its HST centroid (see similar cases in e.g., \citealp{Maiolino2015, Carniani2018}), with a clear elongation toward the S-E (see Fig. \ref{fig:fig3}a).
This [CII] tail is HST-dark, although a faint spike is possibly detected in the HST/WFC3 F160W image (see Fig. \ref{fig:fig1}b).
It is also broadly coincident with the rest-frame optical counterpart (see UltraVista DR4 Ks-band image in Fig. \ref{fig:fig1}c), indicating a UV slope steeply rising towards longer wavelengths and therefore a significant presence of obscuring dust (see e.g., \citealp{Calzetti2000}), as also suggested by the $\gtrsim 4 \sigma$ ALMA detection of co-spatial FIR-continuum emission (see Fig. \ref{fig:fig3}a). 
With the currently available ALMA resolution, 
it is unclear whether this [CII] tail is produced by a dust-obscured star-forming region within $\#A$, 
or it is associated to a dusty neighbour at the same redshift.
%
%
%
\newline

\begin{figure*}
	\centering
	\includegraphics[width=1\textwidth]{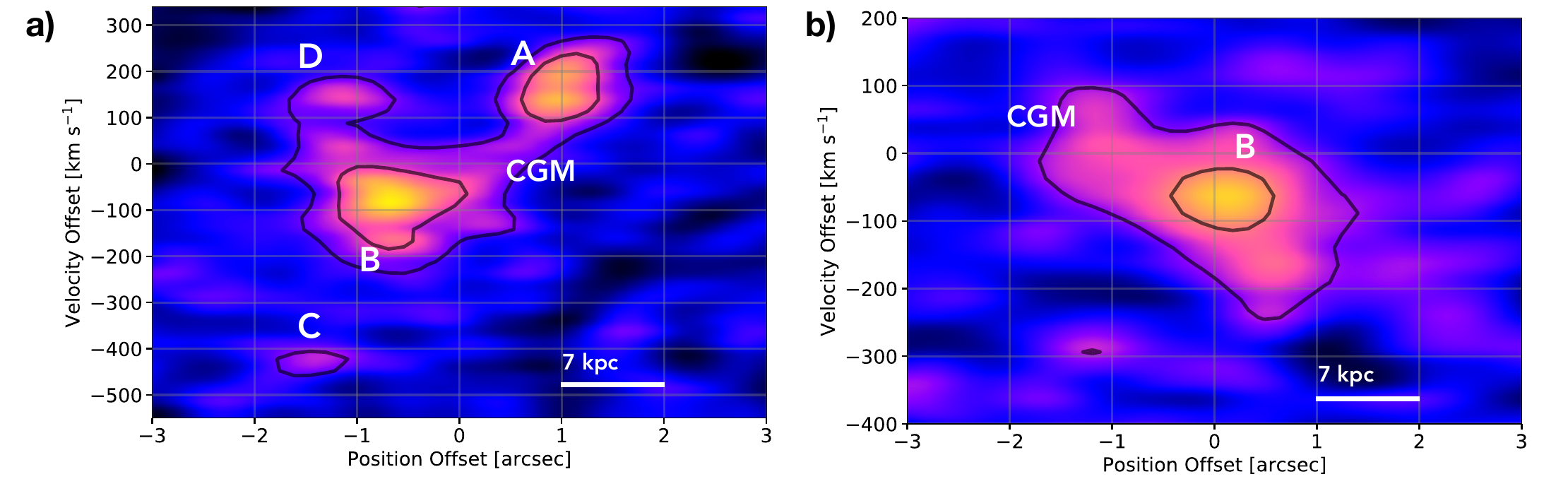}
	\caption{
		{\it (a)} PV diagrams taken along the common axis between the [CII]-flux centroids of $\#A$ - $\#B$, 
		and the orthogonal axis {\it (b)},
		with an  averaging width of five pixels.
		Black contours indicate the $[3, 6] ~ \sigma$ significance levels, where $\sigma_{\rm pv} \sim 0.3$ mJy/beam in both cases.
		A scale of 7 kpc is shown to lower right.
	}
	\label{fig:fig5}
\end{figure*}


The final [CII] spectra of $\#A$ and $\#B$ are shown in Fig. \ref{fig:fig4}a (upper and lower panel, respectively), where we choose an intermediate redshift $z_{\rm ref} = 4.570$ as a common zero-velocity reference for the system%
\footnote{This choice helps when comparing the single galaxy spectra with the emission line profile of the full [CII] emitting system, as discussed further on, whose centroid corresponds to a $z_{\rm ref} = 4.570$.}. 
Galaxy $\#A$, at $z_{\rm [CII]}^{\#A} = 4.573$, has a [CII]-luminosity of $L_{\rm [CII]}^{\#A} = 3.5 ~(\pm 0.3) \times 10^8 ~ {\rm L_\odot}$,
and a ${\rm FWHM}_{\#A} = 190$ km s$^{-1}$,
while $\#B$, at $z_{\rm [CII]}^{\#B} = 4.568$ (blueshifted by $\rm \Delta v_{BA} = -270$ km s$^{-1}$ with respect to $\#A$) 
is broader (${\rm FWHM}_{\#B} = 260$ km s$^{-1}$) and $\sim 2 \times$ more luminous, $L_{\rm [CII]}^{\#B} = 6.6 ~(\pm 0.4) \times 10^8 ~ {\rm L_\odot}$.
Interestingly, the second spectrum (see lower panel of Fig. \ref{fig:fig4}a) shows two additional [CII] emitters, dubbed $\#C$ and $\#D$, spectrally offset by a few hundreds of km s$^{-1}$, and spatially close to $\#B$ 
(less than a $\sim 1''$-beam in projection; see velocity integrated map in the Appendix \ref{sec:appendix2}).
%
The line properties of $\#B$, $\#C$ and $\#D$ (total flux, centroid and FWHM) are estimated from a 3-Gaussian modelling of the spectrum.
Although their integrated flux is detected at $> 4 \sigma$ (see also Appendix \ref{sec:appendix2}), 
$\#C$ and $\#D$ are very faint ($L_{\rm [CII]}^{\#C} = 7.7 ~\pm 1.5 \times 10^7 ~ {\rm L_\odot}$, $L_{\rm [CII]}^{\#D} = 8.4 ~\pm 1.9 \times 10^7 ~ {\rm L_\odot}$),
and narrow (${\rm FWHM}_{\#C} = 40$ km s$^{-1}$, ${\rm FWHM}_{\#D} = 57$ km s$^{-1}$),
and are most likely small satellite galaxies (or gas clumps) orbiting around or accreting onto $\#B$.
\newline

%

%

%

To visualize the total [CII] emission arising from the full system we produce a velocity-integrated flux map by collapsing the spectral channels in the velocity range $[-500 : +500]$ km s$^{-1}$, using a consistent $z_{\rm ref}$ as discussed above.
Such a broad velocity domain is needed to safely include the four detected [CII]-emitting galaxies (see spectra in Fig. \ref{fig:fig4}a).
The resulting velocity-integrated [CII] map is shown in Fig. \ref{fig:fig4}b.
We find a very large [CII]-emitting structure extending on circumgalactic scales, with a projected diameter of $\sim 30$ kpc.
Such a metal-enriched gaseous structure encloses the single galaxy components, and shows a large amount of [CII] arising from regions significantly displaced ($\gtrsim 10$ kpc) from the rest-frame UV / optical counterparts, especially along the direction orthogonal to the $\#A$ - $\#B$ axis.
However, because of its 2D nature, the [CII] flux map shown in Fig. \ref{fig:fig4}b suffers from the dilution of signal emitted by sources with a width much narrower than the velocity-filter adopted for the spectral integration.
Therefore, to better visualize the 3D distribution of the gaseous structure we produce two position-velocity (PV) diagrams, one extracted along the $\#A$ - $\#B$ axis (Fig. \ref{fig:fig5}a), and the other along the orthogonal direction, centred on $\#B$ (Fig. \ref{fig:fig5}b).
Both diagrams highlight the complex morpho-kinematic structure of the system, clearly showing the presence of circumgalactic tails around the interacting galaxies (see e.g., the [CII]-emitting regions labelled as $'$CGM$'$ in both panels of Fig. \ref{fig:fig5}).
%

\begin{figure*}
	\centering
	\includegraphics[width=1\textwidth]{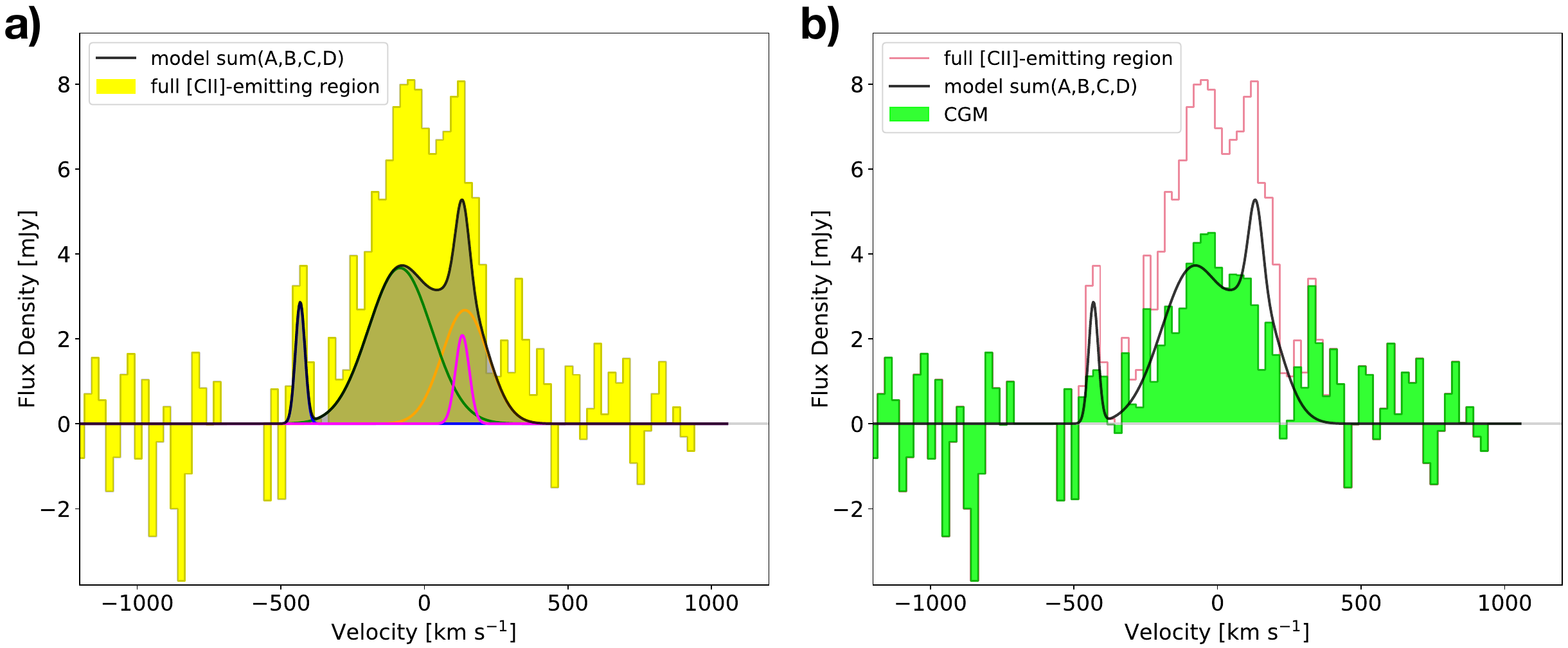}
	\caption{
		The spectrum of the total [CII] emission arising from the full system (Sec. \ref{sec:almaAnalysis}), yellow in {\it (a)} and red in {\it (b)}, is compared with the sum (black shaded area) of the Gaussian models of individual galaxy components (coloured solid lines in {\it (a)}) and with their difference (green spectrum in  {\it (b)}), corresponding to the emission arising from the circumgalactic gaseous envelope.
	}
	\label{fig:fig6}
\end{figure*}

One important piece of information is the net contribution of the four interacting galaxies themselves to the overall [CII] emission arising from the extended gaseous structure.
To distinguish between the relative $'$galaxy$'$ and $'$CGM$'$ [CII] budgets, we extract a spectrum of the total [CII]-emitting system from a region within the 1-$\sigma$ contour of the velocity-integrated map (see grey solid line in Fig. \ref{fig:fig4}b), and compare it with 
the sum of the single $\#A$, $\#B$, $\#C$ and $\#D$ spectra (see Fig. \ref{fig:fig6}a).
We note that, due to the dynamical complexity of our object, 
such a procedure is more accurate than extracting relative budgets directly from the velocity-integrated map, because of the limitations of the latter, as discussed above.
We find that the full [CII]-emitting structure has a total luminosity of $L_{\rm [CII]}^{tot} = 2.3~(\pm 0.2) \times 10^9 ~ {\rm L_\odot}$, i.e., $\sim2$ times larger than the sum of the [CII]-emitting galaxies.
This implies that about $50 ~(\pm6) ~\%$ (see Fig. \ref{fig:fig6}b) of the total [CII] luminosity  resides {\it between} the individual components, $L_{\rm [CII]}^{CGM} = 1.1 ~(\pm 0.2) \times 10^9 ~ {\rm L_\odot}$,
in a sort of circumgalactic and metal-rich gaseous {\it envelope} around the merging galaxies.
This fraction of [CII] emission arising from the CGM is larger than what typically found around the few individual normal isolated galaxies (i.e., with not evident signs of major/minor mergers) that show a [CII]-halo (see \citealp{Fujimoto2020}), suggesting a link with the merging-nature of our system.

In the next section we discuss possible mechanisms responsible for such large amount of circumgalactic [CII] emission.

\section{Discussion: gas  mixing in the CGM}\label{sec:discussion}

\subsection{Dynamical interactions and shocks}
A natural scenario to explain the origin of the luminous gaseous envelope is that the circumgalactic [CII] traces gas stripped during gravitational encounters occurring in the dense merging system.
Strong dynamical interactions between the four galaxies might therefore remove a significant amount of interstellar gas from their outer regions, polluting the CGM with chemically enriched material, 
and producing the complex morphology (see Fig. \ref{fig:fig3} and Fig. \ref{fig:fig4}a) and kinematics (see Fig. \ref{fig:fig5}) observed.

In this scenario, our observations could be a high-$z$ analogue of local studies of circumgalactic stripped [CII] in local groups.
For instance \cite{Appleton2013} detect a broad [CII] emission arising from an extended shocked filament ($\sim 35$ kpc) produced by supersonic collisions between galaxies in the compact local group Stephan's Quintet.
They observe very large [CII]/Polycyclic Aromatic Hydrocarbon (PAH) and  [CII]/FIR ratios, which cannot be accounted for by [CII] emission originating from photo-dissociation regions (PDRs; \citealp{Stacey1991, Stacey2010}) only, and suggest that circumgalactic [CII] can be excited by the dissipation of mechanical energy (turbulence and shocks) injected by strong dynamical interactions.
Such a mechanism is expected to be more frequent and more disruptive at high-$z$
because of 
(i) higher merger rates (e.g., \citealp{Conselice2006,Wetzel2009,Lotz2011})
and (ii) generally higher gas fractions (e.g., \citealp{Santini2014, Dessauges2017, Schinnerer2016, Scoville2017,Tacconi2018}),
and might also explain previous similar observations of extended [CII] emission around multi-component systems at $z\sim5$ (see ${\rm HZ6}$ and ${\rm HZ8}$ in \citealp{Capak2015, Faisst2017}).
We will further explore this hypothesis by extending similar morpho-spectral decomposition analyses presented in this paper to other major/minor merging systems in the ALPINE survey (see a discussion on ALPINE mergers in \citealp{LeFevre2020} and \citealp{Jones2020}) in future works.

Interestingly, 
since as discussed in Sec. \ref{sec:protocluster} our merging system is close to an overdensity peak of the proto-cluster ${\rm PCl J1001+0220}$ at $z\sim4.57$ (see \citealp{lemaux2018}), 
the merger-induced CGM pollution might be 
a natural channel to feed and enrich the nascent proto- intra-cluster-medium 
(ICM; see e.g., \citealp{Cucciati2014}, \citealp{Tozzi2003}, and a recent review by  \citealp{Overzier2016}).
Thus, our results could be some of the first observational evidence at high redshift
that mergers and gas stripping are key  mechanisms in driving the build-up and the chemical evolution of the ICM,
as predicted by models and numerical simulations (see e.g., \citealp{Gnedin1998,Toniazzo2001, Tornatore2004, DeLucia2004, Kapferer2007}).
Also, the peculiar location of our system within a large-scale overdense structure suggests that it sits in proximity of a cosmic web node where several filaments from the IGM might be feeding into it. 
Shocks are predicted to occur as the intergalactic gas accretes onto massive galaxies and mixes
 with their CGM (see e.g., \citealp{Birnboim2003, Keres2009}), and they likely contribute to the [CII] excitation.
To quantitatively assess this scenario we will need new deep observations of the large-scale environment tracing the hydrogen gas, representing the bulk of the intergalactic streams, that can be traced e.g., through Ly$\alpha$ with MUSE (which has a field of view of $\sim 400$ kpc at this redshift).
\newline

\subsection{Galactic outflows}

Another possible contributing scenario is that a fraction of the circumgalactic metal-enriched gas traced by the extended [CII] envelope is expelled from the internal ISM of galaxies through galactic outflows.
Evidence of star formation-driven winds has been recently observed at $z>4$ (see \citealp{Sugahara2019,Faisst2019,Ginolfi2020,Cassata2020}),
and repeated episodes of outflows are thought to be responsible for the origin of diffuse [CII]-{\it halos} ($\sim 20$ kpc) detected around isolated massive star-forming high-$z$ galaxies (see \citealp{Faisst2017, Fujimoto2019, Ginolfi2020, Pizzati2020, Fujimoto2020}).
However, a rough estimate suggests that outflows might not be the dominant mechanism for injecting the large amount of circumgalactic cold gas observed in our system (see Fig. \ref{fig:fig4}b and Fig. \ref{fig:fig6}).
Using the $L_{\rm [CII]}-M_{\rm gas}$ relation calibrated by \cite{Zanella2018} and the [CII] luminosity arising from the CGM (i.e., about $50\%$ of the total $L_{\rm [CII]}$ (see last paragraph of Sec. \ref{sec:almaAnalysis}) we obtain a rough estimate of the dense mass of in the [CII] gaseous envelope, $M_{\rm gas}^{\rm CGM} \sim  3\times 10^{10} ~{\rm M_\odot}$.
Then, using typical mass outflow rates of $\dot{M}_{\rm out} \sim 30 ~ {\rm M_\odot ~ yr^{-1}}$ (as measured by \citealp{Ginolfi2020} in the sub-sample of highly star-forming ALPINE galaxies), and assuming that both $\#A$ and $\#B$ significantly contribute to gas-expulsion%
\footnote{
We assume that the contribution from the satellite galaxies $\#C$ and $\#D$ to gas-pollution through outflows is negligible. Indeed, from their $L_{\rm [CII]}$, using the $L_{\rm [CII]} - {\rm SFR}$ relation calibrated by \citealp{Schaerer2020}, we find ${\rm SFRs}  \lesssim 10 ~{\rm M_\odot ~ yr^{-1}}$, and galaxies with such a low SFR at high-$z$ have been reported to not show signatures of prominent outflows (e.g., \citealp{Ginolfi2020}), nor signs of significant amount of circumgalactic emission (e.g., \citealp{Ginolfi2020, Fujimoto2020}).
}%
, we find that outflows would require at least $\sim$ 0.6 Gyr to expel the estimated amount of dense gas in the CGM.
We note that this estimate is very conservative, since it is derived under the unlikely assumptions that (i) re-accretion does not occur, and (ii) both $\#A$ and $\#B$ have constantly had during their past growth a sufficient SFR to produce powerful outflows.
\newline

%

\subsection{Small faint satellites}

A third complementary process would be that part of the extended [CII] emission is induced by the star formation of several small satellites, which are unresolved by our ALMA observations and faint in the rest-frame UV HST images (the few light spots around $\#A$ and $\#B$ in the HST maps, see Fig. \ref{fig:fig1}a and Fig. \ref{fig:fig1}b, may belong to this category).
To estimate the efficiency of this mechanism we estimate the possible diffuse SFR produced by faint satellites
by integrating the rest-frame UV light within an aperture defined as the 1$\sigma$ level of the [CII] full structure (see grey line in Fig. \ref{fig:fig1}a and Fig. \ref{fig:fig4}b), and then subtracting the UV flux arising from regions defined as the blue ellipses (see Sec. \ref{sec:almaAnalysis} and Fig. \ref{fig:fig4}b) associated with the individual galaxy components $\#A$, $\#B$, $\#C$ and $\#D$. 
%
Using a standard \cite{Kennicutt2012} calibration, we estimate a total diffuse UV-based ${\rm SFR}_{\tiny{CGM}}^{\tiny{UV}}= 13 \pm 5 ~{\rm M_\odot ~ yr^{-1}}$,
%
which is about one order of magnitude lower than expected assuming a $L_{\rm [CII]} - {\rm SFR}$ relation%
\footnote{We used three different calibrations, which give consistent results: 
(i) ${\rm SFR}_{\tiny{CGM}}^{\tiny{[CII]}} \sim 129 \pm 46 ~{\rm M_\odot ~ yr^{-1}}$ from \cite{Schaerer2020}, calibrated for galaxies at $4<z<6$;
(ii) ${\rm SFR}_{\tiny{CGM}}^{\tiny{[CII]}} \sim 124 \pm 40 ~{\rm M_\odot ~ yr^{-1}}$ from \cite{DeLooze2014}, calibrated for local galaxies;
(iii) ${\rm SFR}_{\tiny{CGM}}^{\tiny{[CII]}} \sim 125 ~{\rm M_\odot ~ yr^{-1}}$ from the model of \cite{Lagache2018}.
Errors on the SFR are computed using the uncertainties on the parameters of the relations.
}
(see e.g., \citealp{DeLooze2014, Lagache2018, Schaerer2020}; see a similar calculation in \citealp{Fujimoto2019}), ${\rm SFR}_{\tiny{CGM}}^{\tiny{[CII]}} \sim 125~ {\rm M_\odot ~ yr^{-1}}$.
This suggests that, 
unless the diffuse SFR was exceptionally dust-obscured, which is disfavoured by the non-detection of ALMA FIR-continuum
(see also \citealp{Fudamoto2020}, showing that low-mass objects are not expected to have high $L_{\rm IR} / L_{\rm UV}$), 
small faint satellites do not significantly account for the detected large circumgalactic [CII] emission.
\newline
\newline
Altogether, gas stripping by dynamical interactions and (with a possibly relative lower efficiency) galactic outflows and small faint satellites
jointly contribute to the gas mixing and the metal enrichment of the CGM around high-$z$ merging systems.
%
To resolve these mechanisms and quantitatively assess their relative efficiencies, 
we urge (i) deeper/ higher-resolution ALMA FIR-lines mapping and 
(ii) future rest-frame optical emission line observations with JWST,
to be interpreted with tailored zoom-in cosmological simulations (e.g., \citealp{Kohandel2019, Pallottini2019, Graziani2020}).

\section{Conclusions}\label{sec:conclusions}

In this work we present ALMA observations of [CII] and FIR-continuum emission in/around a merging system at $z\sim4.57$, 
observed as a part of the ALPINE survey (see \citealp{LeFevre2020,Faisst2019,Bethermin2020}).
Combining information from a rich set of panchromatic ancillary data (including imaging and spectroscopy from rest-frame UV to the near-IR; \citealp{Faisst2019}) 
with a morpho-spectral decomposition of the ALMA data (see Sec. \ref{sec:almaAnalysis}), 
we find the results summarised below.

\begin{itemize}

\item $\rm VC\_9780$ (dubbed $\#A$) and  $\rm C15\_705574$ (dubbed $\#B$), 
with star formation rates ${\rm SFR}_{\#A} = 38^{+29}_{-14} ~ {\rm M_\odot ~ yr^{-1}}$ and ${\rm SFR}_{\#B} = 106^{+9}_{-65} ~ {\rm M_\odot ~ yr^{-1}}$ respectively, 
and similar stellar masses of $M_{\star} \sim 10^{10} ~ {\rm M_\odot}$,
are undergoing a major merger (stellar mass ratio $r_{\rm mass} \sim 0.9$; see Sec. \ref{sec:ancillary}), at a close projected separation of $\sim 13$ kpc (see Fig. \ref{fig:fig1}), with a small velocity offset of $\Delta v \sim 270 ~ {\rm km~s^{-1}}$.
ALMA data reveal the presence of two faint and narrow (${\rm FWHM} \sim 50 ~ {\rm km~s^{-1}}$) [CII]-emitting satellites (see Fig. \ref{fig:fig4}) close to $\#B$, and most likely in the process of orbiting around (or accreting into) it.
We find that our merging system belongs to the proto-cluster 
${\rm PCl J1001+0220}$ (\citealp{lemaux2018}; see Sec. \ref{sec:protocluster}), and it is coincident to one of its overdensity peaks (see Fig. \ref{fig:fig2}).
\newline

\item We produce a velocity-integrated map collapsing the broad velocity range $\rm [-500 : +500] ~ {\rm km~s^{-1}}$
to visualize the total [CII] emission arising from the full system,
and find a very large structure extending over circumgalactic scales, out to a diameter-scale of $\sim$ 30 kpc (see Fig. \ref{fig:fig4}), surrounding the galaxy components.
A fraction of [CII] flux appears significantly displaced ($\gtrsim 10$ kpc) from the rest-frame UV counterparts of $\#A$ and $\#B$.
Both flux maps (see Fig. \ref{fig:fig3}, Fig. \ref{fig:fig4}) and PV diagrams (see Fig. \ref{fig:fig5}) indicate a disturbed morphology and complex kinematics of the overall gas structure.
\newline

\item We compare the flux arising from the extended [CII]-emitting region with the sum of [CII] fluxes emitted by single galaxies, 
and find that about $50 ~\%$ of the total emission arises from a gaseous envelope distributed between the individual components of the system (see Fig. \ref{fig:fig6}; see also discussion about possibly similar objects in \citealp{Faisst2017}).
While a contribution of processed material expelled by galactic outflows (see e.g., \citealp{Ginolfi2020, Fujimoto2020})
and SFR-induced [CII]  emitted by small faint satellites
should be not negligible, 
we argue that most of the detected circumgalactic emission originates from the effect of 
gas stripping induced by strong gravitational interactions (see a discussion in Sec. \ref{sec:discussion}), in analogy with observations of tidal tails of shock-excited [CII] in local compact groups (e.g., \citealp{Appleton2013}). 
This might also represent a natural channel to feed and enrich the ICM in the large-scale proto-cluster environment surrounding our system (see Sec. \ref{sec:protocluster} and Fig. \ref{fig:fig2}).
\newline

\end{itemize}

Altogether our findings suggest that dynamical interactions at high-$z$ can be an efficient mechanism for extracting gas out of galaxies and mixing the CGM with chemically evolved material.
Deeper and higher resolution ALMA data, as well as future JWST rest-frame optical emission line observations, will be necessary to study more in details the key role of mergers in the baryon cycle of distant galaxies.

	
\section*{Acknowledgements}

The authors would like to thank the anonymous referee for her/his useful suggestions.
This paper is based on data obtained with the ALMA Observatory, under Large Program 2017.1.00428.L. 
ALMA is a partnership of ESO (representing its member states), NSF (USA) and NINS (Japan), together with NRC (Canada), MOST and ASIAA (Taiwan), and KASI (Republic of Korea), in cooperation with the Republic of Chile. 
The Joint ALMA Observatory is operated by ESO, AUI/NRAO and NAOJ. 
G.C.J.and R.M. acknowledge the ERC Advanced Grant 695671 "QUENCH'' and support by the Science and Technology Facilities Council (STFC).
C.G. and M.T. acknowledge the support from  a grant PRIN MIUR 2017.
S.C. acknowledges support from the ERC Advanced Grant INTERSTELLAR H2020/740120.
The Cosmic Dawn Center is funded by the Danish National Research Foundation under grant No. 140.
Some of the material presented in this paper is based upon work supported by the National Science Foundation under Grant No. 1908422.
This program receives financial support from the French CNRS-INSU Programme National Cosmologie et Galaxies.

This paper is dedicated to the memory of Olivier Le F\`evre, PI of the ALPINE survey.

\appendix  
\section{Description  of the VMC mapping  technique}\label{sec:appendix1}

Here we briefly describe the VMC mapping technique used to produce Fig. \ref{fig:fig2}, discussed in Sec. \ref{sec:protocluster}. 
We refer to \cite{lemaux2018} (and reference therein) for more details.

To reconstruct the density field in a given redshift range, a large number of Monte Carlo iterations of the input spectroscopic and photometric catalogs are run. 
For each Monte Carlo realization, we begin with all objects that have spectroscopic information, sampling a random uniform distribution for each object. 
For a given object, if the sample exceeds the value of a predetermined reliability flag for that redshift, we throw away the spectral redshift and, for that Monte Carlo realization, rely on the photometric redshift information. 
If the uniform sampling does not exceed the reliability threshold, we retain the spectral information for that object in that realization. 
For those objects where the spectral information was disavowed along with those objects that had no spectral information then have their photometric redshifts
tweaked by sampling a reconstruction of their full $P(z)$. 
Thus, at the end of a given iteration, all objects in the combined spectral and photometric 
catalog have a set redshift. We then set a redshift boundary to generate the two-dimensional map and Voronoi tessellate over all objects
whose redshifts, spectral or photometric, fall within the redshift boundaries. The local density value, $\Sigma_{VMC}$, of a given object for a given iteration is 
the inverse of the area of the Voronoi cell surrounding it. This process is repeated for the set number of Monte Carlo realizations, with the density values of each
realization sampled by a standardized grid of 75$\times$75 $h_{70}^{-1}$ proper kpc, and the pixel values from all realizations are then median combined. 
As in other works, we rely here on local overdensity rather than local density as the former has been found to be less sensitive to observational effects. 
The local overdensity value for each grid point is then computed as $\log(1+\delta_{gal}) \equiv \log(1+ (\Sigma_{VMC}-\tilde{\Sigma}_{VMC})/\tilde{\Sigma}_{VMC})$, 
where $\tilde{\Sigma}_{VMC}$ is the median $\Sigma_{VMC}$ for all grid points over which the map was defined, i.e., excluding an $\sim1$ arcmin wide border 
region to mitigate edge effects.

\section{[CII] flux maps of $\#C$ and $\#D$}\label{sec:appendix2}

\begin{figure*}
	\centering
	\includegraphics[width=1\textwidth]{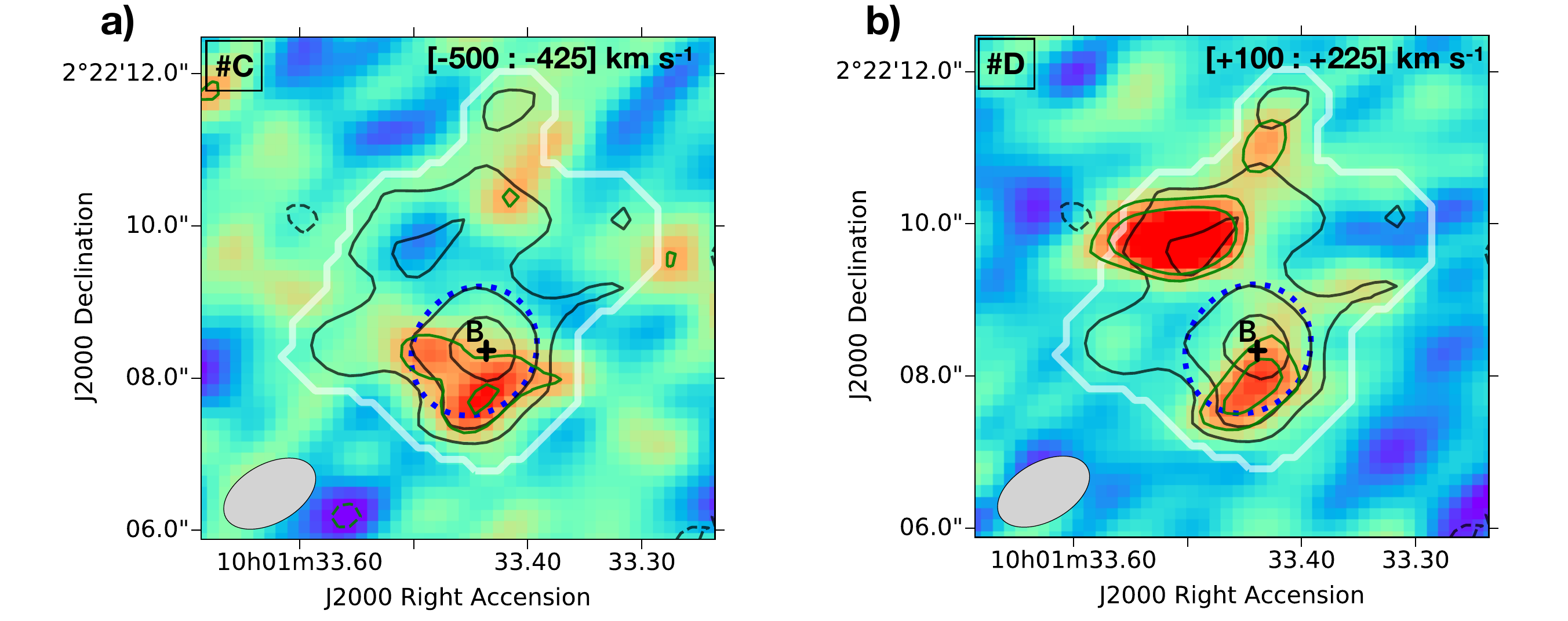}
	\caption{
{\it (a)}  Velocity-integrated [CII] maps of $\#C$ and  {\it (b)} $\#D$, obtained by collapsing 
the spectral channels corresponding to a $2\times{\rm FWHM}$ interval centred on their $z_{\rm [CII]}$ 
(see spectra in Fig. \ref{fig:fig4}a, and values in Table 1).
The velocity ranges used to collapse the datacubes are shown in the top-left corners.
The rest-frame UV centroid of $\#B$ is indicated with a plus symbol.
The dashed blue ellipse defined in Fig. \ref{fig:fig3} (see Sec. \ref{sec:almaAnalysis}) is shown for reference.
The grey solid line and the black contours indicate the significance level of the [CII] flux emitted by the full system as defined in 
Fig. \ref{fig:fig4}b.
The green solid (dashed) contours indicate the positive (negative) significance levels at $[3,4]\sigma$ of the [CII] flux maps of $\#C$ {\it (a)} and $\#D$ {\it (b)}. 
To help the visualization of {\it (b)}, 
we used a maximum value of surface brightness in the colour-map range as defined by excluding the bright spot at North (corresponding to a fraction of the emission of $\#A$).
The ALMA beam size is given in the bottom-left corner. North is up and east is to the left.
	}
	\label{fig:figAppendix}
\end{figure*}

\bibliographystyle{aa}
\bibliography{biblio.bib}

\end{document}